\documentclass[reprint,amsmath,amssymb,aps,prl,]{revtex4-2}

\usepackage{graphicx}
\usepackage{dcolumn}
\usepackage{bm}

\begin{document}

\title{Two Nambu--Goldstone zero modes for rotating Bose--Einstein condensates}

\author{Rodney E. S. Polkinghorne}
\author{Tapio P. Simula}
\affiliation{Optical Sciences Centre, Swinburne University of Technology, Melbourne 3122, Australia}

\begin{abstract}
We consider rotating finite size vortex arrays in Bose--Einstein condensates that are confined by cylindrically symmetric external potentials. We show that such systems possess two exact Nambu--Goldstone zero modes associated with two spontaneously broken continuous symmetries of the system. We verify our analytical result via direct numerical diagonalizations of the Bogoliubov--de Gennes equations. We conclude by comparing rotating vortex lattices in superfluids to supersolids and discrete time crystals.
\end{abstract}

\maketitle

For Lorentz invariant quantum field theories the Goldstone theorem posits that for every spontaneously broken continuous symmetry, there should exist a corresponding massless Nambu--Goldstone boson \cite{Nambu1960a,Goldstone1961a}. Under certain circumstances, the Higgs mechanism allows circumventing this whereby the would-be massless particle may acquire a finite mass by coupling to the Higgs field \cite{Englert1964a,Higgs1964a,Guralnik1964a,HiggsCMS,HiggsATLAS}.

In low energy condensed matter systems in the absence of exact Lorentz invariance the theory carries over with the massless bosons corresponding to gapless quasiparticle Nambu--Goldstone zero modes \cite{Beekman2019a}. Prominent examples include phonons in solids and in superfluids, and spin waves in magnetic systems. Conventional superconductors are characterised by an excitation energy gap and the absence of a zero mode associated with the superconducting order parameter in that case is attributed to the Anderson--Higgs mechanism \cite{Anderson1963a}. The Higgs amplitude mode, a condensed matter counterpart to the Higgs boson, has recently been observed in cold atom experiments \cite{Endres2012a,Pekker2015a}. Classification and counting rules for the number of expected zero modes in effective field theories of low energy condensed matter systems have been established that explain how linear dependence between the generators of the broken symmetries may lead to redundancies, reducing the total number of zero modes with respect to the number of spontaneously broken continuous symmetries \cite{Ueda2006a,Uchino2010a,Nitta2015a,Watanabe2013a,Hidaka2013a,Hidaka2020a}.

An interacting scalar Bose--Einstein condensate (BEC) in its ground state is described by a complex valued order parameter $\phi({\bf r})=|\phi({\bf r})|e^{iS({\bf r})}$ with a constant real valued spatial phase function $S({\bf r})$. The birth of a BEC is associated with a spontaneous breaking of the continuous U(1) symmetry as the atoms become phase locked, and the BEC wavefunction $\phi({\bf r})$ is the resulting Nambu--Goldstone zero mode. The condensate has a chemical potential $\mu$ that causes the condensate phase to continuously sample all U(1) phases according to $\phi({\bf r},t)=\phi({\bf r})e^{-i\mu t/\hbar}$. As such, the effect of the zero mode is to restore the broken symmetry in an average sense by `rotating' the ground state phase so that all possible broken symmetry phases are sampled equitably over time. 

When such a BEC is \emph{spatially} rotating, quantised vortices nucleate in the condensate, and these localised point-like particles spontaneously arrange into a regular pattern breaking the continuous SO(2) rotation symmetry. In equilibrium, a triangular vortex lattice is typically realised \cite{Madison2000a,AboShaeer2001a,Haljan2001a,Hodby2001a,Zwierlein2005a}. The vortex lattice ground state of a rotating BEC spontaneously breaks two continuous symmetries and, according to the Goldstone theorem it would be reasonable to anticipate two Nambu--Goldstone zero modes. Nevertheless, it has been suggested that out of the two phonons only one would survive in the thermodynamic limit due to the aforementioned redundancy \cite{Watanabe2013a}.

When a single off-centre vortex is present in a trapped BEC, the vortex orbits around the trap centre with a constant angular frequency \cite{Anderson2000a,Freilich2010a}. Similarly, a vortex lattice in a laboratory frame rotates as a rigid body at an orbital angular frequency $\Omega=\kappa n_v/2$, where $n_v$ is the two-dimensional vortex density and $\kappa$ is the quantum of circulation. These systems thus respond to the broken rotation symmetry by a rotating excitation that in a time-averaged sense restores the broken continuous symmetry. However, the low-energy vortex mode \cite{Coddington2003a,Schweikhard2004a,Baym2003a,Simula2004a,Mizushima2004a,Baksmaty2004a,Cozzini2004a,Bhattacherjee2004a,Sonin2005a,Simula2010a,Matveenko2011a,Simula2013a} associated with this symmetry breaking has remained elusive.

Here we show that rotating Bose--Einstein condensates indeed do possess an \emph{exact} zero energy Kelvin--Tkachenko Bogoliubov quasiparticle vortex mode associated with the SO(2) symmetry breaking, in addition to the condensate zero mode associated with the U(1) symmetry breaking. We begin by proving analytically the existence of these two zero modes. We then explicitly demonstrate their presence via direct numerical diagonalizations of the Bogoliubov--deGennes equations. Next we rationalize our findings in the context of previous analyses, and finally we discuss the implications of the obtained result and its relation to the supersolid phases of matter and rapidly rotating condensates. 

Let us consider a stationary order parameter $\phi({\bf r})$ that satisfies the Gross--Pitaevskii equation (GPE) in the reference frame rotating at angular frequency $\Omega$. The GPE can be expressed as 
\begin{equation}
    (T+V({\bf r}) -\mu -\Omega L_z)\phi({\bf r}) = -g|\phi({\bf r})|^2\phi({\bf r}),
\end{equation}
where $T=-\hbar^2\nabla^2/ (2m)$ is the kinetic energy operator, $V({\bf r})$ is external trap potential here assumed to be cylindrically symmetric,  $L_z=-i\hbar\left(x{\partial\over\partial y}-y{\partial\over\partial x}\right)$ is the angular momentum operator normal to the plane of rotation, and $g$ is the coupling constant \cite{Dalfovo1999a}. The state is normalised according to $\int|\phi({\bf r})| d{{\bf r}}=N$, where $N$ is the number of atoms.

The elementary excitations of such a system contain co- and counter-rotating terms, and perturb the ground state according to \cite{Dalfovo1999a}
\begin{equation}
\psi({\bf r}, t) = [\phi({\bf r})+\epsilon e^{-i\omega_q t}u_q({\bf r})-\epsilon^*e^{i\omega_q t}v_q^*({\bf r})]e^{-i\mu t/\hbar},
\label{bogpert}
\end{equation}
where $\epsilon$ is a complex number with an infinitesimal magnitude.
In order for $\psi({\bf r}, t)$ to satisfy the time-dependent Gross-Pitaevskii equation, the mode functions must satisfy the Bogoliubov--de Gennes (BdG) eigenvalue equations \cite{Dalfovo1999a}
\begin{equation}
    \begin{pmatrix}
        \mathcal{L}   &\mathcal{D}_{12} \\
        \mathcal{D}_{21}    &-\mathcal{L}^\ast
    \end{pmatrix}
    \begin{pmatrix}u_q({\bf r})\cr v_q({\bf r})\end{pmatrix} =
    E_q\begin{pmatrix}u_q({\bf r})\cr v_q({\bf r})\end{pmatrix}
\label{bdg}
\end{equation}
for the quasiparticle amplitudes $u_q({\bf r})$ and $v_q({\bf r})$ corresponding to the eigenenergies $E_q=\hbar \omega_q$, where $q$ uniquely labels the quantum states, which satisfy orthonormalisation condition $\int (u^*_iu_j - v^*_iv_j) d{\bf r}=\delta_{i,j}$. For the scalar BEC the matrix elements are $\mathcal{L} =T+V({\bf r}) -\mu -\Omega L_z +2 g|\phi({\bf r})|^2$ and $\mathcal{D}_{12} = -\mathcal{D}_{21}^\ast=-g\phi({\bf r})^2$, such that the BdG eigenvalues may have non-zero imaginary components as in `non-Hermitian' quantum mechanics.

\begin{figure}
    \centering
    \includegraphics[width=\hsize]{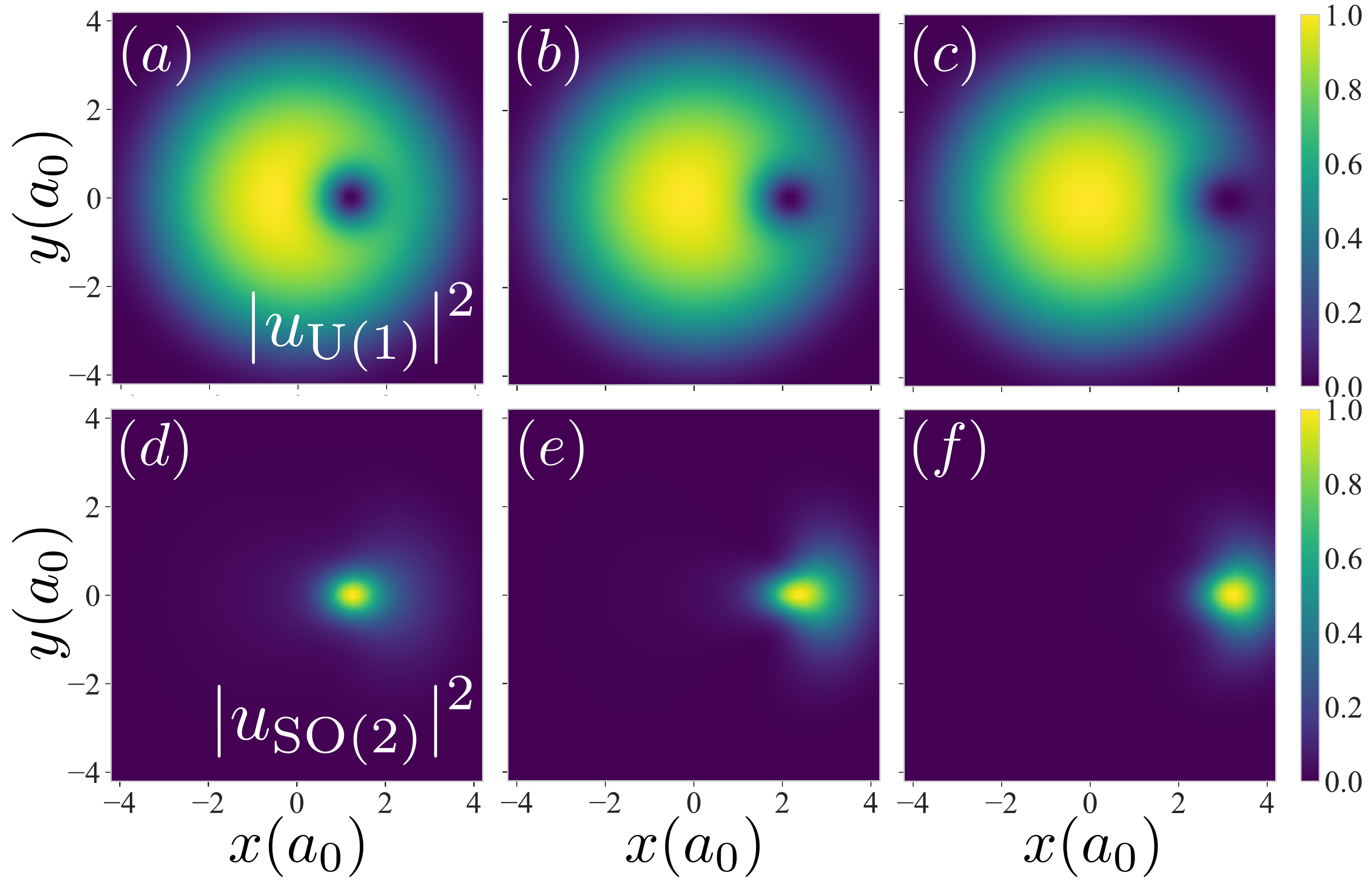}
    \caption{The two Nambu--Goldstone zero modes for a single off-centered vortex in a Bose--Einstein condensate. Densities of the condensate quasiparticle mode $|u_{\rm U(1)}|^2=|\phi|^2$ that restores the $\rm U(1)$ symmetry is shown in (a)-(c) for three different stationary states for frame rotation frequencies $\Omega =(0.355,0.398,0.479)\;\omega_\perp$, respectively. The corresponding densities of the Kelvin quasiparticle mode $|u_{\rm SO(2)}|^2=|L_z\phi|^2$ that restores the $\rm SO(2)$ symmetry is shown in (d)-(f). The color scale is normalised to the peak density.}
    \label{fig:i}
\end{figure}

Suppose then that the system has a symmetry generated by the operator $P$.  Physically, there must exist an `excitation' corresponding to an infinitesimal transformation $\psi({\bf r},t)=(1+i\epsilon'P)\phi({\bf r},t)$ with real valued $\epsilon'$, and $\omega_q=0$. Setting $u_q=v_q^*=P\phi({\bf r})$, and $\epsilon={i\over 2}\epsilon'$ results in $i\epsilon'P\phi({\bf r})=\epsilon u-\epsilon^*v^*$, consistent with Eq.~(\ref{bogpert}).

The BdG equations have an exact zero energy, $E_1=0$, solution 
\begin{equation}
u_1({\bf r})=v^*_1({\bf r})=\phi({\bf r}),\label{phizero}
\end{equation}
which is straightforward to verify by direct substitution to Eq.~(\ref{bdg}). This condensate mode is the well known Nambu--Goldstone zero mode associated with the U(1) symmetry breaking, generated by $P=1$. In addition, we have found that the BdG equations have another exact zero energy, $E_2=0$, solution
\begin{equation}
u_2({\bf r})=v^*_2({\bf r})=L_z\phi({\bf r}),
\label{zero}
\end{equation}
associated with the SO(2) symmetry breaking, with generator $P=L_z$. 

To show that Eq.~(\ref{zero}) is a solution of Eq.~(\ref{bdg}), we first note that the auxliary operator $A=\mathcal{L}-2 g|\phi({\bf r})|^2 $ and $L_z$ commute, $[A,L_z]=[A^\ast,L_z]=0$, and that $A\phi = - g|\phi|^2\phi$ and $(L_z\phi)^\ast=-L_z\phi^\ast$. Therefore, direct substitution of the putative zero mode solution, Eq.~(\ref{zero}), to the BdG equations yields
\begin{align}
&\begin{pmatrix}
        A+2g|\phi({\bf r})|^2&-g\phi({\bf r})^2\cr
        g\phi({\bf r})^{\ast 2}&-A^\ast-2g|\phi({\bf r})|^2
    \end{pmatrix}L_z
    \begin{pmatrix}\phi({\bf r})\cr -\phi({\bf r})^\ast\end{pmatrix}  \\  
=&\begin{pmatrix}
        AL_z\phi +2g|\phi|^2L_z\phi +g\phi^2 L_z\phi^\ast \cr
        g\phi^{\ast 2} L_z\phi + A^\ast L_z\phi^\ast +2g|\phi|^2L_z\phi^\ast  
    \end{pmatrix} \notag \\ 
=&    g\begin{pmatrix}
        -L_z\left(|\phi|^2\phi\right)+2|\phi|^2(L_z\phi)+\phi^2(L_z\phi^\ast)\cr
         \phi^{\ast 2}(L_z\phi)-L_z\left(|\phi|^2\phi^\ast\right)+2|\phi|^2(L_z\phi^\ast)
    \end{pmatrix}=0.\notag 
\end{align}
This proof survives self-consistency condition accounting for the presence of quantum fluctuations and thermal atoms, and generalises to continuous symmetries generated by a generic operator $P$, provided that $P$ is a derivation.

\begin{figure}
    \centering
    \includegraphics[width=0.98\columnwidth]{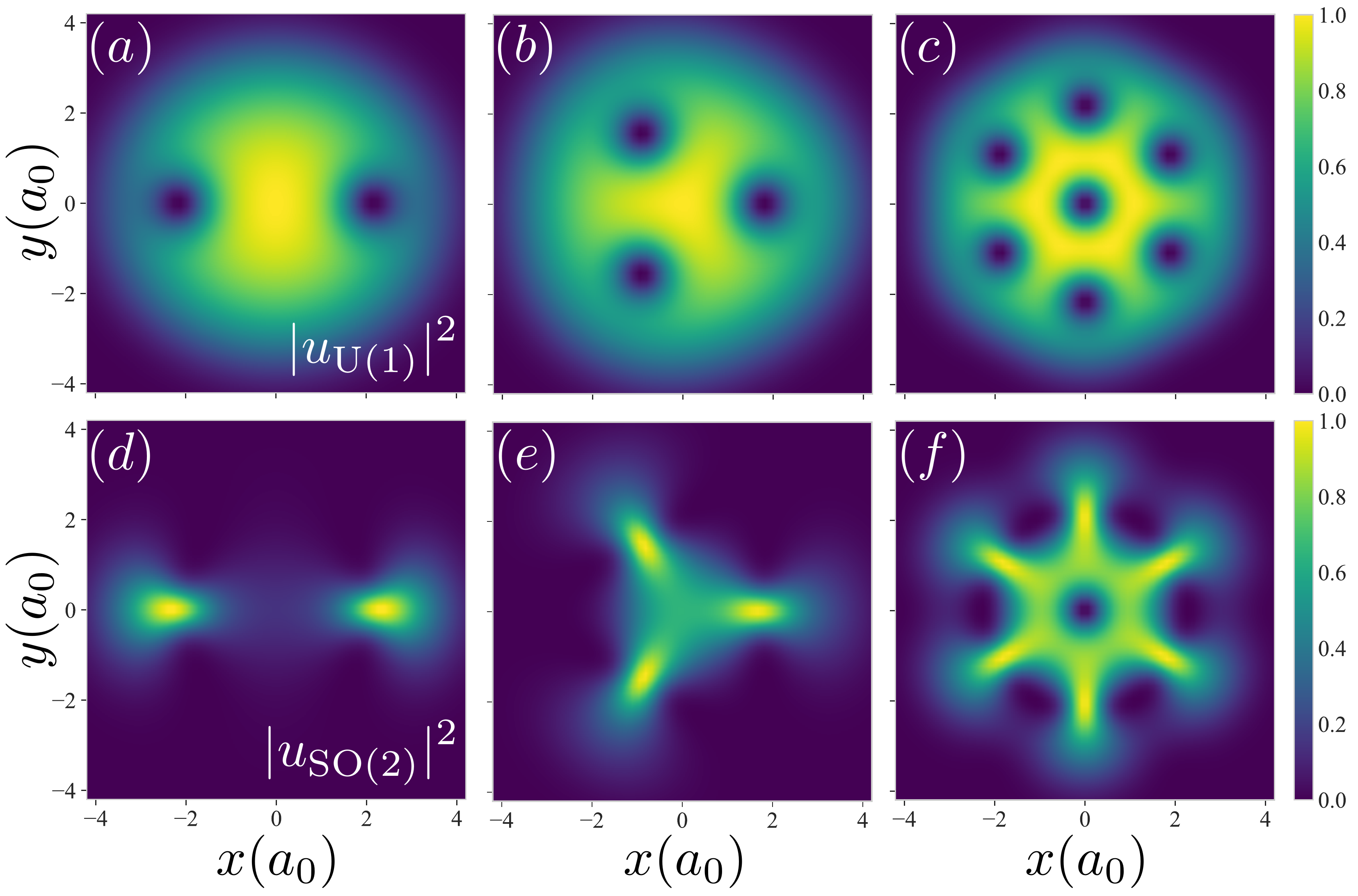}
    \caption{The two Nambu--Goldstone zero modes for 2, 3, and 7 vortex arrays in a Bose--Einstein condensate rotating at respective orbital angular frequencies of $\Omega =(0.482,0.606,0.800)\;\omega_\perp$. Notation is as in Fig.~\ref{fig:i}.}
    \label{fig:ii}
\end{figure}

We note that in Eq.~(\ref{bogpert}) a real part of $\epsilon$ corresponds to two counter rotating terms that cancel each other. On their own, each of these terms would shift the order parameter in a way canonically conjugate to the symmetry generator. Adding a real multiple of the condensate mode shifts the condensate particle number, while adding a real multiple of the $L_z\phi$ zero mode translates the vortices radially, shifting the angular momentum. These kind of perturbations would thus correspond to Higgs amplitude modes in this system. The atom removal method to excite the Tkachenko mode \cite{Coddington2003a} may be viewed from this perspective.

To verify that the Eq.~(\ref{zero}) indeed appears as a true zero mode in the elementary excitation spectrum of rotating Bose--Einstein condensates, we have performed direct numerical diagonalizations of the BdG equations for a range of stationary states. Following the standard protocols, a stationary state solution of the Gross--Pitaevskii equation is first found in the rotating reference frame. The obtained condensate wavefunction determines the pair potential in the BdG equation that is then diagonalised to yield the quasiparticle eigenstates of the system. Our two-dimensional numerical calculations are conducted using Julia programming language \cite{Polkinghorne2021a}. The dimensionless coupling constant $g_{\rm 2D}N/(\hbar\omega_\perp a^2_0)=100$, where $a_0=\sqrt{\hbar/m\omega_\perp}$ is the harmonic oscillator frequency and $g_{\rm 2D}$ is the effective two-dimensional coupling constant.

We first revisit the single vortex case due to its direct relevance to the rotational symmetry breaking and the problem of vortex nucleation \cite{Isoshima1999a,Caradoc1999a,Raman2001a,Penckwitt2002a,Simula2002a,Isoshima2003a,Ueda2006a,Dagnino2009a}. Figure \ref{fig:i} shows the densities $|u_q|^2$ for the two zero modes $E_1=E_2=0$ corresponding to the U(1) symmetry for which $|u_1|^2=|v_1|^2=|\phi |^2$ (top row), and the SO(2) symmetry for which $|u_2|^2=|v_2|^2=|L_z\phi |^2$ (bottom row) for the case of a single off-centered vortex whose stationary radial position is set by the frame rotation frequency $\Omega$. This SO(2) Kelvin mode \cite{Dodd1997a,Isoshima1997a,Virtanen2001a,Bretin2003a,Fetter2004a,Simula2008a} has a significant density in the vortex core where the U(1) condensate mode density vanishes. When the vortex is about to denucleate at the condensate edge, the Kelvin zero mode hybridizes with the surface mode that mediates the symmetry breaking topological quantum phase transition between the vortex and non-vortex states, associated with the closing of a gap in the quasiparticle excitation spectrum.  

To demonstrate that both zero modes are present for all symmetry broken states irrespective of the vortex number, Fig.~\ref{fig:ii} shows the densities $|u_q|^2$ of the two zero modes for the case of small arrays of 2, 3, and 7 vortices. Similarly to the single vortex case, the SO(2) Kelvin--Tkachenko zero mode density has maxima at the cores of the off-center vortices, highlighting the spatial crystalline order of the vortex array. As in the single vortex case, the condition $|u_q|^2=|v_q|^2$ is satisfied for all the zero modes making it straightforward to identify them also by their quasiparticle amplitudes.

\begin{figure*}
    \centering
    \includegraphics[width=1.9\columnwidth]{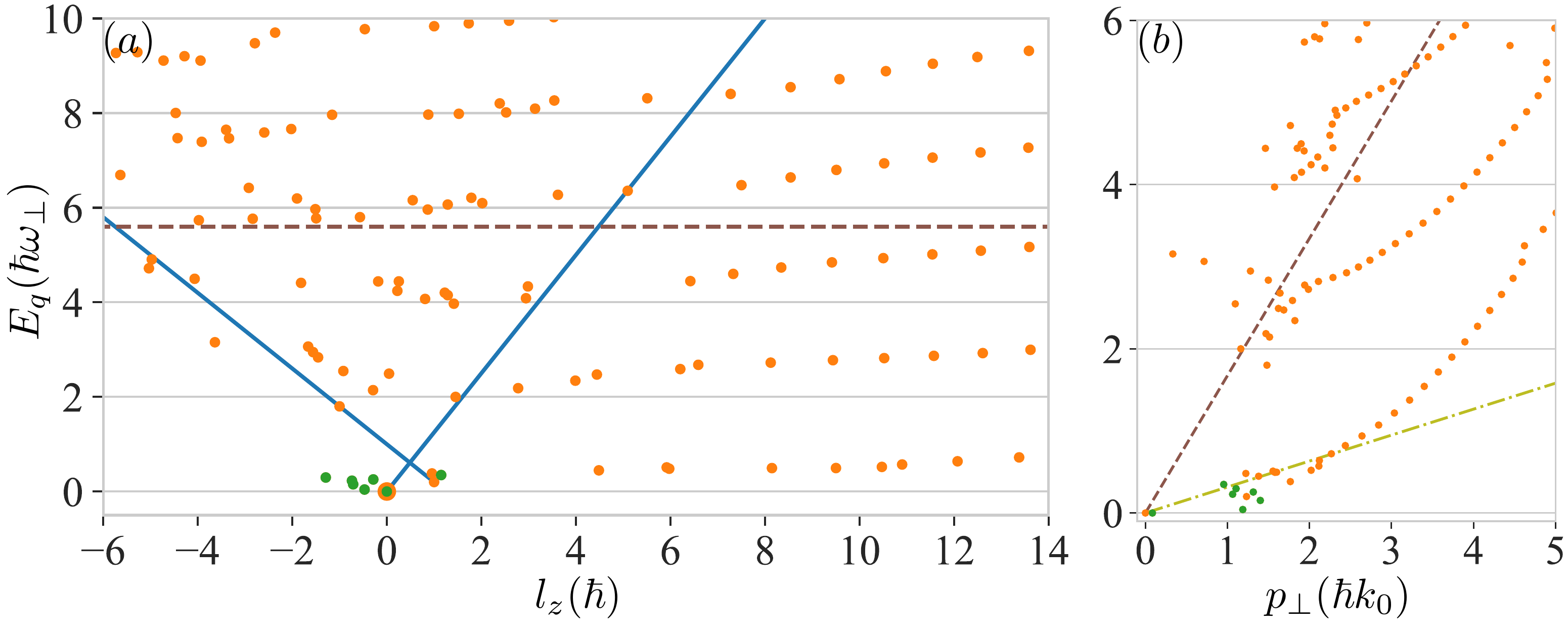}
    \caption{Elementary excitation energy spectra as functions of angular (a) and linear (b) momenta for a rotating Bose--Einstein condensate with seven vortices. In (a) the straight lines have a slopes $-\Omega$ and $1/\Omega$ and the dashed horizontal line is the chemical potential $\mu$. The seven Kelvin--Tkachenko vortex modes are highlighted with green markers. In (b) the dashed line has a slope $c_s=\sqrt{\mu/2m}$ and the dashdotted line has a slope $c_{\rm T}=\sqrt{\hbar\Omega/8m}$.}
    \label{fig:iii}
\end{figure*}

Having confirmed the presence of the zero modes, it is instructive to place them in the context of the overall structure of the elementary excitations. Figure \ref{fig:iii} (a) shows the quasiparticle excitation spectrum for the seven vortex array as a function of quasiparticle angular momentum per particle. The Landau levels of non-interacting harmonic oscillator, whose level spacing equals the cyclotron gap $2\hbar\omega_\perp$, are provided for reference (gray horizontal lines). The magnitude of the chemical potential $\mu=5.6\;\hbar\omega_\perp$ is shown using the dashed line. For this case the parameter $\Gamma_{\rm LLL}=\mu/(2\hbar\Omega) = 3.5$, and the system is not far from the mean-field quantum Hall regime \cite{Schweikhard2004a}.

The blue lines illustrate the frame rotation at frequency $\Omega = 0.8\,\omega_\perp$. Consequently, the two Kohn modes are shifted to $E_{-1}=1.8\,\omega_\perp$ and $E_{+1}=0.2\,\omega_\perp$ such that the line passing through these modes has the slope $-\Omega$ \cite{Zambelli1998a,Chevy2000a}. The line orthogonal to the one intersecting the Kohn modes has slope $1/\Omega$ and passes through the origin and the breathing mode, which due to the ${\rm SO}(2,1)$ hidden symmetry has a frequency of $2\omega_\perp$ \cite{Pitaevskii1997a}. We obtain this value within numerical uncertainty, such that a presence of a quantum anomaly \cite{Olshanii2010a,Peppler2018a,Holten2018a} seems unlikely in this system. 

The two overlapping zero modes are shown in Fig.~\ref{fig:iii} (a) using the larger green/orange marker. The remaining six Kelvin--Tkachenko vortex modes (for $N_v$ vortices the spectrum contains $N_v$ vortex quasiparticle modes) are shown using green markers. The lowest Landau level (LLL) is comprised of the $N_v$ vortex modes together with the low energy surface modes. The Alfv\'en wave of the vortex plasma, corresponding to the inertial wave in the rotating superfluid, is gapped and in the limit $\Omega=\omega_\perp$ will oscillate at the cyclotron frequency $2\Omega$. By contrast, the U(1) and SO(2) phonons are gapless terminating at their respective zero modes.

Figure~\ref{fig:iii} (b) shows the quasiparticle energies as a function of their momentum $p_\perp(q)=\sqrt{|K_q(u)+K_q(v)-2K_1(u)|}$, where $K_q(w)=\langle w_q|T|w_q\rangle/\langle w_q|w_q\rangle$. As in (a) the vortex modes are highlighted with green markers. The dashed line has a slope $c_s =b\sqrt{\mu/m}$ and the dashdotted line has a slope $c_{\rm T} =bl_{B}\Omega/2$ \cite{Tkachenko1966a,Fetter1967a} where $l_B=\sqrt{\hbar/m\Omega}$ is the `magnetic' length. The speed ratio of these two sounds is $c_s/c_{\rm T}=\sqrt{8\Gamma_{\rm LLL}}$. For this system $c_s/c_{\rm T}\approx 5.3$ and we have used $b=1/\sqrt{2}$ in Fig.~\ref{fig:iii} (b).

Previous numerical calculations have studied the low energy excitations of vortex lattices by performing either time-dependent GPE simulations \cite{Simula2004a,Mizushima2004a} or by solving the BdG equations for 2D \cite{Mizushima2004a,Baksmaty2004a} and 3D \cite{Simula2010a,Simula2013a,Simula2013b} systems. Despite being in reasonable agreement with the experiments \cite{Coddington2003a,Schweikhard2004a}, these previous numerical works have not identified the two zero modes, apart from the fortuitous exception of \cite{Simula2013a}. The reason for this may be the numerical complexity of diagonalizing exceedingly large matrices, a problem that has often been solved using iterative methods. Guided by our analytical result, Eq.~(\ref{zero}), it is straightforward to calculate the order parameter of the SO(2) zero mode by using the GPE solution, to confidently identify its presence also in the BdG spectrum.

The $(n=1,m=0)$ Tkachenko mode, where the integers $n$ and $m$ denote the number of radial and azimuthal nodes, was observed to have very low oscillation frequency, approaching zero in the rapid rotation limit \cite{Coddington2003a,Schweikhard2004a}, and theoretical continuum models \cite{Tkachenko1966a,Fetter1967a,Baym2003a,Cozzini2004a,Bhattacherjee2004a,Sonin2005a,Matveenko2011a,Watanabe2013a} predicted this mode to be either linearly or quadratically dispersing soft mode. In this context, it is noteworthy that the $(1,0)$ Tkachenko mode is not the lowest energy quasiparticle excitation mode in these systems. The exact $(0,0)$ zero mode has no radial or azimuthal nodes, however, the `motion' of the vortices generated by this mode is deceivingly similar to that of the $(1,0)$ mode.

As pointed out in Ref.~\cite{Watanabe2013a}, it is interesting to draw parallels between vortex lattices and supersolid states of matter \cite{Tanzi2019a,Natale2019a,Guo2019a,Norcia2021a}. A key characteristic of a supersolid is the presence of multiplicity of broken continuous symmetries. Specifically, a supersolid simultaneously possess diagonal long-range order (spatial crystal) and off-diagonal long-range order (superfluid). The observable signature of this dual symmetry broken supersolid phase is the presence of at least two phonon modes, one corresponding to the sound wave of the superfluid and one corresponding to the phonon of the crystal vibrations. In a vortex lattice the superfluid order enables the propagation of phonons as density waves \cite{Andrews1997a,Engels2005a} and the presence of a vortex crystal results in the propagation of Kelvin--Tkachenko vortex waves \cite{Coddington2003a,Schweikhard2004a}.

A usual two-dimensional solid state crystal has three broken continuous symmetries, two for translations and one for rotation. However, linear dependence between the generators of these symmetries results in redundancies leaving the system with reduced number of zero modes \cite{Watanabe2013a}. In a trapped BEC, all translation symmetries are already explicitly broken by the confining potential and the discrete translation invariance within the vortex lattice does not amount to additional Nambu--Goldstone zero modes. However, the rotation symmetry does remain unbroken in the non-rotating ground state and therefore the emergence of the vortex lattice spontaneously breaks the continuous SO(2) rotation symmetry, resulting in the emergence of the second Nambu--Goldstone zero mode. 

The presence of two-fold ground state degeneracy is also interesting from the perspective of discrete time crystals \cite{Sacha2018a,Else2020a}. The spontaneously emerging six fold discrete rotation symmetry of the vortex lattice means that the state is recurrent in the laboratory reference frame with a period $T_6= T_\Omega/6$, where $T_\Omega=2\pi/\Omega\approx 4\pi/(\kappa n_v)$ is the natural rigid body rotation period of the lattice. The vortex lattice is an excited state in the absence of external driving, yet it is protected from tunnelling to the non-rotating ground state by the conservation of angular momentum if the external potential is cylindrically symmetric. In practice, in low temperature experiments that have good control of the trap asymmetry, the vortex lattice is a metastable state with an unmeasurably long life time in comparison to the life time of the host superfluid.

In conclusion, we have shown that rotating vortex lattices in scalar Bose--Einstein condensates have two exact Nambu--Goldstone zero modes in their quasiparticle excitation spectra. How do these gapless modes give way for gapped strongly correlated quantum fluids deep in the LLL \cite{Schweikhard2004a,Fletcher2021a,Mukherjee2021a} is a fascinating contemporary open question.

\begin{acknowledgements}
We thank Martin Zwierlein and his research group for stimulating discussions. This work was performed on the OzSTAR national facility at Swinburne University of Technology. The OzSTAR program receives funding in part from the Astronomy National Collaborative Research Infrastructure Strategy (NCRIS) allocation provided by the Australian Government.
This research was funded by the Australian Government through the Australian Research Council (ARC) Discovery Project DP170104180 and the Future Fellowship FT180100020.
\end{acknowledgements}

\bibliography{apssamp}

\end{document}